\documentclass{aa}
\newcommand{\bit}{\begin{itemize}}
\newcommand{\eit}{\end{itemize}}
\newcommand{\ben}{\begin{enumerate}}
\newcommand{\een}{\end{enumerate}}
\newcommand{\bde}{\begin{description}}
\newcommand{\ede}{\end{description}}

\usepackage{graphicx}
\usepackage{lscape}

\begin{document}
\title{Metal enrichment of the intra-cluster medium over a Hubble time for merging and relaxed galaxy clusters}
\author{W. Kapferer$^1$,
        T. Kronberger$^1$,
        J. Weratschnig$^1$,
        S. Schindler$^1$,
        W. Domainko$^2$,
        E. van Kampen$^1$,
        S. Kimeswenger$^1$,
        M. Mair$^1$,
        M. Ruffert$^3$}

\institute{ $^1$Institut f\"ur Astro- und Teilchenphysik,
            Universit\"at Innsbruck,
            Technikerstr. 25,
            A-6020 Innsbruck, Austria \\
            $^2$Max-Planck-Institut f\"ur Kernphysik,
            Saupfercheckweg 1,
            69117 Heidelberg, Germany\\
            $^3$School of Mathematics,
            University of Edinburgh,
            Edinburgh EH9 3JZ,
            Scotland, UK }

\offprints{W. Kapferer, \email{Wolfgang.E.Kapferer@uibk.ac.at}}

\date{-/-}

\abstract{We investigate the efficiency of galactic mass loss,
triggered by ram-pressure stripping and galactic winds of cluster
galaxies, on the chemical enrichment of the intra-cluster medium
(ICM). We combine N-body and hydrodynamic simulations with a
semi-numerical galaxy formation model. By including simultaneously
different enrichment processes, namely ram-pressure stripping and
galactic winds, in galaxy-cluster simulations, we are able to
reproduce the observed metal distribution in the ICM. We find that
the mass loss by galactic winds in the redshift regime z$>$2 is
$\sim$10\% to 20\% of the total galactic wind mass loss, whereas the
mass loss by ram-pressure stripping in the same epoch is up to 5\%
of the total ram-pressure stripping mass loss over the whole
simulation time. In the cluster formation epochs z$<$2 ram-pressure
stripping becomes more dominant than galactic winds. We discuss the
non-correlation between the evolution of the mean metallicity of
galaxy clusters and the galactic mass losses. For comparison with
observations we present two dimensional maps of the ICM quantities
and radial metallicity profiles. The shape of the observed profiles
is well reproduced by the simulations in the case of merging
systems. In the case of cool-core clusters the slope of the observed
profiles are reproduced by the simulation at radii below $\sim$300
kpc, whereas at larger radii the observed profiles are shallower. We
confirm the inhomogeneous metal distribution in the ICM found in
observations. To study the robustness of our results, we investigate
two different descriptions for the enrichment process interaction.

\keywords{Hydrodynamics -- Methods: numerical -- Galaxies:
abundances -- Galaxies: interactions -- Galaxies: general --
intergalactic medium -- Galaxies: evolution}}

\authorrunning {W. Kapferer et al.}
\titlerunning {Metal enrichment of the intra-cluster medium over a Hubble time}

\maketitle
%

\section{Introduction}

From numerous X-ray observations (Fukazawa et al. 1998, Schmidt et
al. 2002, Furusho et al. 2003, Sanders et al. 2004, Fukazawa et al.
2004, Hayakawa et al. 2004, Pratt et al. 2006, De Grandi et al. 2004
and Gitti \& Schindler 2004) it is well known that the intra-cluster
medium (hereafter ICM) harbours on average $\sim$0.6\% heavy
elements in mass, i.e. 0.3 in solar abundance. By using state of the
art X-ray space telescopes it was found that the metals are not
homogenously distributed within the ICM. Metallicity profiles show a
relatively flat distribution in non-cool core clusters, while there
is an increase in the central metallicity in cool core clusters (De
Grandi et al. 2004, Vikhlinin et al. 2005,
Pratt et al. 2006).\\
Many groups have derived recently detailed two dimensional
metallicity maps (e.g. Sanders et al. 2004, Durret et al. 2005,
O'Sullivan et al. 2005, Sauvageot et al. 2005, Werner et al. 2006,
Sanders \& Fabian 2006, Hayakawa et al. 2006, Finoguenov et al.
2006, Bagchi et al. 2006). All these maps show an inhomogeneous
distribution of the heavy elements with several maxima, complex
metal patterns and a non-spherically symmetric distribution.\\
Many processes were investigated to explain the origin of these
metals, i.e. galactic winds (De Young 1978), ram-pressure stripping
(Gunn \& Gott 1972), galaxy-galaxy interactions (Gnedin 1998,
Kapferer et al. 2005), active galactic nuclei (AGNs)(Moll et al.
2006) and intra-cluster supernovae (Domainko et al.
\cite{domainko04}). Ram-pressure stripping, for instance, has been
found in several Virgo cluster galaxies by HI observations (Cayatte
et al. 1990; Veilleux et al. 1999, Vollmer et al. 1999, Vollmer
2003, Kenney et al. 2004, Vollmer et al. 2004a,b; Koopmann \& Kenney
2004, Crowl et al. 2005, Levy et al. 2006). Galactic winds, driven
by supernova explosion within cluster galaxies have been observed
and studied in detail in many systems in multiple wavelengths (see
reviews by Heckman et al. 2003 and Veilleux et al. 2005). The mass
fraction of metals in these outflows depends on galaxy parameters
and on environmental conditions. Martin (1999) found that the mass
loss by galactic winds scales linearly with the galaxies star
formation rate. It was shown, that in the centre of massive clusters
the pressure of the ICM can suppress galactic winds (Schindler et
al. 2005).  Also hydrodynamic simulations of outflows have been
performed (Tenorio-Tagle \& Munoz-Tunon 1998, Strickland
\& Stevens 2000).\\
Many groups have investigated the chemical evolution of the ICM
using numerical simulations. De Lucia et al. (2004) and Nagashima et
al. (2005) studied the enrichment of the ICM with N-body simulations
combined with a semi-analytic galaxy formation model. Their
conclusion is that nearly all metals found in the ICM are ejected by
supernova-driven galactic winds by the most massive galaxies.
Additionally they claim, that there is a light chemical evolution
since z = 1. So far, none of these processes alone was sufficient to
produce the observed pollution of the ICM. Therefore recent
simulations were not able to reproduce the observed amount and the
spatial distribution of metals in the ICM by galactic winds
(Tornatore et al. 2004, Kapferer et al. 2006) or ram-pressure
stripping (Domainko et al. 2006). Whereas the effects of galactic
winds (e.g. Tornatore et al. 2004, Scannapieco et al. 2006 and Romeo
et al. 2006) or ram-pressure stripping (e.g. Domainko et al. 2006)
on galaxies have been studied analytically and numerically by many
authors, we focus mainly on the evolution and distribution of the
metallicity in the ICM. In this paper we investigate the impact on
the enrichment of the ICM by ram-pressure stripping and galactic
winds of cluster galaxies. Compared to our previous work (Schindler
et al. 2005, Kapferer et al. 2006, Domainko et al. 2006, Moll et al.
2006) we extend this investigation to the epoch where the first
galaxies form and compare the efficiency of galactic winds and
ram-pressure stripping over a Hubble time.

\section{Simulations}
\subsection{Numerical methods}

As described in Schindler et al. (2005), Kapferer et al. (2006) and
Domainko et al. (2006) we apply different code modules to calculate
appropriately each of the galaxy-cluster components. The
cosmological model used has the following parameters: $\Omega_{\rm
m}=0.3$, $\Omega_\Lambda=0.7$, $h=0.7$, $\sigma_8=0.93$,
$\Omega_{\rm b}=0.02 h^{-2}$.

\subsubsection{Dark Matter}
The non-baryonic component is calculated using GADGET2 (Springel
2005) as an N-body code, with constrained random fields as initial
conditions (Hoffman \& Ribak (1991), implemented by van de Weygaert
\& Bertschinger (1996)). The N-body code provides the underlying
evolution of the dark matter (DM) potential for the hydrodynamic
code and the dynamically fully described orbits of model cluster
galaxies.

\subsubsection{Galaxy formation model}

The properties of the galaxies are calculated by an improved version
of the galaxy formation code of van Kampen et al. (1999). The galaxy
formation model is semi-numerical in the sense that the merging
history of galaxy haloes is taken directly from the cosmological
N-body simulation. Halo mergers are identified in the N-body
simulations as events in which a new halo is formed that is both
virialised and contains at least two progenitors. We distinguish
between major mergers, where two progenitors have at least 30 \% of
the mass of the merger remnant, and minor mergers, which we treat as
accretion events. During any merger event, all galaxies involved
undergo a starburst whose intensity depends on the progenitor/merger
remnant mass ratio (small for minor mergers), and whose duration
depends on the relaxation time of the new halo (short in the case of
minor mergers). Galaxy-galaxy mergers are treated differently: all
galaxies within a dark matter halo or subhalo suffer from dynamical
friction, and will merge with the central galaxy eventually. The
dynamical friction time scale is reset after a major merger event
(assuming that orbits will be randomized by the event), but not
after a minor merger event. When a satellite galaxy merges with the
central galaxy of a halo or subhalo, another starburst will occur,
using up most of the cold gas left in the satellite. Not all is
turned into stars due to feedback, however. In between merger events
star formation is quiescent, and we therefore have two types of star
formation modes: quiescent star formation in disks with a threshold
according to the Kennicutt criterion, and starbursts resulting from
major merger events, including halo-halo and galaxy-galaxy mergers.
Gas that cools within virialised haloes is assumed to settle in a
disk with an exponential profile. We use the models of Mo, Mao \&
White (1998) to set the disk-scale length, with the distinction that
we measure the angular momentum of the dark halo from the N-body
simulation data instead of assuming an analytical model for that.
Stellar evolution is modelled using the stellar population synthesis
models of Bruzual \& Charlot (2003). The metallicity of the stellar
populations affects the cold and hot gas components of a galaxy: a
fraction of the metals formed in stars is ejected by SNe and stellar
winds into the surrounding interstellar medium (ISM). The metals
that end up in the hot gas component also affect the cooling rate,
which depends on metallicity (Sutherland \& Dopita 1993). Chemical
evolution is modelled as in  Matteucci \& Francois (1989), which
includes SN types I and II. The evolution of the metals is followed
for the stellar populations as well as for the cold and hot gas
reservoirs, and exchanges are also tracked. Material returned to the
cold ISM by stellar winds and supernovae has been chemically
enriched by the nuclear processes inside the stars.

\subsubsection{Hydrodynamics}

For the treatment of the ICM we use a comoving hydrodynamic code
with shock capturing scheme (PPM, Collela \& Woodward 1984), with a
fixed mesh refinement scheme (Ruffert 1992) on four levels and
radiative cooling (Sutherland \& Dopita (1993)). The major
improvement with respect to the previously used setup is the usage
of comoving coordinates, which was implemented in the hydrodynamic
code by applying a operator splitting method. This approach allows
us to apply cosmological initial conditions for the hydrodynamic
simulation. The largest grid covers a comoving volume of (20
Mpc)$^3$. Each finer grid covers $\frac{1}{8}$ of the volume of the
next larger grid. With a resolution of 128$^3$ grid cells on each
grid we obtain a finest resolution of ($\sim$19.5 kpc)$^3$ comoving
for each cell on the innermost grid. The N-body  and hydrodynamics
code are calculated starting at $z=40$, while the semi-numerical
galaxy formation covers the redshift interval from $z=20$ to $z=0$.
\\
\noindent This setup allows us to address questions like:
\begin{itemize}
\item{what is the efficiency of ram-pressure stripping versus galactic winds as a function of redshift?}
\item{what is the spatial distribution of metals in the ICM ejected by galactic winds and/or lost by ram-pressure stripping?}
\end{itemize}

\subsection{Calculation of the mass loss due to galactic winds}

The mass loss due to galactic winds is calculated using the star
formation rate (SFR) ($M_{\star}$ [$M_{sun}$/yr]) of a given galaxy,
as proposed by Martin (1999). We use the same approach as Springel
\& Hernquist (2003), who found that a mass loss due to galactic
winds $\dot{M}=\eta M_{\star}$, with a constant fraction $\eta$, is
reasonable to fit observations (see Springel \& Hernquist (2003) and
references therein). The same was found by Tornatore et al. (2004)
who used $\eta=2$. This heuristic approach gives us the mass loss
for a given spiral galaxy, which we then add to the hydrodynamic
simulation at the galaxy's position. The metals are mixed with the
ICM, which is already present in a computational cell. The
metallicity of the ejected gas is taken from the semi-numerical
galaxy model, which calculates the metals in the hot and cold gas
phase by
\begin{eqnarray}
\dot{m_c} &=& Z_h \dot{M}_{c}-Z_c(1+ \beta -R) \Psi_{SF}+y
\Psi_{SF}\\
\dot{m_h} &=& Z_h \dot{M}_{c}+Z_c \beta \Psi_{SF}+Z_{Prim}
\dot{M}_{new},
\end{eqnarray}
\noindent where $Z_c=m_c/M_c$ ($Z_c$ the metallcity, $m_c$ the mass
of heavy elements and $M_c$ the mass of the cold gas), $Z_h=m_h/M_h$
($Z_h$ the metallcity, $m_h$ the mass of heavy elements and $M_h$
the mass of the hot gas), and $\dot{M}_{new}$ is the accretion of
new baryons by the halo, $\Psi_{SF}$ is the star formation rate,
$Z_{prim}$ the primordial metallicity and $\beta$ the fraction of
supernovae in a stellar population. This results in a lower limit
for the metallicity of the galactic-wind ejecta.

\subsection{Calculation of the mass loss due to ram-pressure stripping}

To calculate the mass loss of cluster galaxies due to ram-pressure
stripping we follow the classical Gunn \& Gott (1972) criterion. In
this model, gas of galactic discs is stripped outside a certain
stripping radius where the ram pressure exceeds the restoring
gravitational forces. According to Vollmer et al. (2001) the Gun \&
Gott (1972) criterion extended for the inclination angle can explain
the HI deficiency of galaxies observed in the region of the Virgo
cluster. We use the same approach as presented in detail in Domainko
et al. (2006) for sub- and supersonic galaxies interacting with the
ICM. The main parameters in the ram-pressure stripping routine are
the relative velocity $v$ of a galaxy with respect to the ICM, the
density $\rho_{ICM}$ of the ICM and the surface density of both, the
stellar $M_{star}$ and the gas $M_{gas}$ mass in the model galaxy.
As shown in Domainko et al. (2006) a certain stripping radius for
each galaxy, depending on the properties of the ICM and the galaxy,
can be derived
\begin{equation}
R_{strip}=\frac{R_0}{2} \times \ln \left( \frac{G M_{star}
M_{gas}}{v^2 \rho_{ICM} 2 \pi R_{0}^{4}}\right),
\end{equation}

\noindent where $R_0$ is the disc scale length of the gas
distribution in the galaxy. Gaseous matter which is located outside
this stripping radius is lost by the galaxy. As a consequence of the
used PPM scheme in the grid-based hydrodynamic calculations, shocks
induced by merger events and their propagation throughout the
cluster are well treated. These shock features have direct impact on
the ram-pressure stripping.

\subsection{Combination of galactic winds and ram-pressure stripping}

We combine both enrichment processes in a multi-step approach. First
we select from the complete sample of model galaxies the spiral
galaxies by requiring a significantly large (i.e. disk scale length
larger than bulge radius) star forming, gaseous disk. In a second
step the star formation rate in the disk is used to calculate the
galactic wind mass loss as described in Sect. 2.2. In order to
investigate the relative importance of both processes we study two
extreme cases. First we consider a strong influence of ram-pressure
stripping on the galactic winds. As long as the galaxy has not
fulfilled the criterion for ram-pressure stripping the galaxy is
allowed to lose gas by galactic winds. When the galaxy loses matter
due to ram-pressure stripping, it will not be affected by galactic
winds anymore. This avoids an overestimation of the mass loss by
galactic winds, especially in the low redshift regime, i.e. z$<$1.5
where galaxy clusters form and dynamical enrichment processes like
ram-pressure stripping become dominant. Numerous observational
results point at a quenching of the star formation in galaxies,
encountering the dense environment of the ICM (e.g. Haines et al.
2006, Gavazzi et al. 2002). This description of the combined
processes leads to a lower limit in the metal enrichment efficiency
for galactic winds at lower redshifts. As a second approach we do
not consider any influence of ram-pressure stripping on the galactic
winds. A galaxy can have a galactic wind and ram-pressure stripping
at the same time. Note that hereafter the index $a$ denotes the
first and $b$ denotes the second combination of the two enrichment
processes.

\section{The properties of the model galaxy clusters}

In order to investigate the efficiency of different enrichment
processes we take a non-cooling flow cluster from a sample of
simulated merging systems, which is a lower mass cluster with
several strong sub-cluster mergers. The second model cluster is a
more massive merging system with nearly more than twice the mass and
several minor merger events.

\begin{itemize}

\item \textbf{Model Cluster A:} The cluster forms  at z$\sim$1.5 and has two major
merger events at z=0.8 and z=0.5. The final total mass is
$1.5\times10^{14}$ M$_{\odot}$ in a sphere of radius 1 Mpc. We show
the X-ray surface brightness and X-ray weighted temperature map at
different redshifts in Fig. \ref{xsb_maps602} and Fig.
\ref{xt_maps}, respectively.

\item \textbf{Model Cluster B:} The formation redshift for this
cluster is z$\sim$1.7. It shows four minor merger events a z=1.4,
z=1.1, z=0.5 and z=0.3. The cluster has a final mass of
3.4$\times10^{14}$ M$_{\odot}$ within a sphere of radius 1
Mpc . We show the X-ray surface brightness and X-ray weighted
temperature map at different redshifts in Fig. \ref{xsb_maps604} and
Fig. \ref{xt_maps_604}, respectively.

\end{itemize}

\section{Results}
\subsection{X-ray weighted maps}

\begin{figure}
\begin{center}
{\includegraphics[width=\columnwidth]{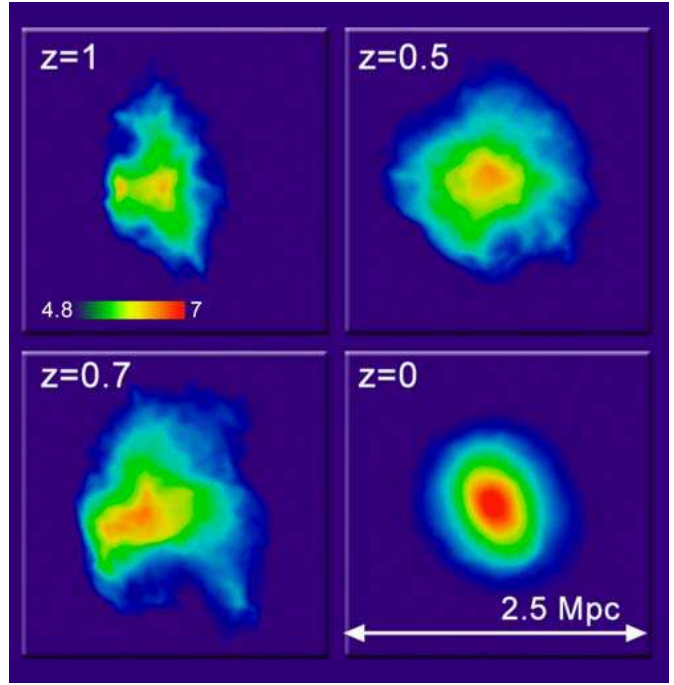}} \caption{X-ray
surface brightness maps for different redshifts (z=1, 0.7, 0.5 and
0) for model cluster A. The maps have 2.5 Mpc comoving on a side and
give the X-ray emission in a logarithmic scale.} \label{xsb_maps602}
\end{center}
\end{figure}

In order to compare the model clusters with observations we extract
X-ray surface brightness, X-ray weighted temperature and metal maps
at different redshifts. In Fig. \ref{xsb_maps602} X-ray surface
brightness maps for model cluster A at redshifts z=1, 0.7, 0.5 and 0
are presented on a logarithmic scale. The area covers (2.5 Mpc)$^2$
in comoving space. The final X-ray surface brightness map has a drop
of $\sim 2$ magnitudes from the cluster centre to r$_{200}$, which
is consistent with observations of non-cooling flow clusters (Arnaud
et al. 2002, De Filippis et al. 2003). The hierarchical formation of
the cluster is clearly visible, especially at redshift 1, when the
cluster is forming. At lower redshifts model cluster A shows an
asymmetric shape, which is a common feature for non cool-core
clusters, that typically undergo several mergers. The model cluster
begins to relax at redshift z=0, nevertheless the spherical shape
has not fully established yet.

\begin{figure}
\begin{center}
{\includegraphics[width=\columnwidth]{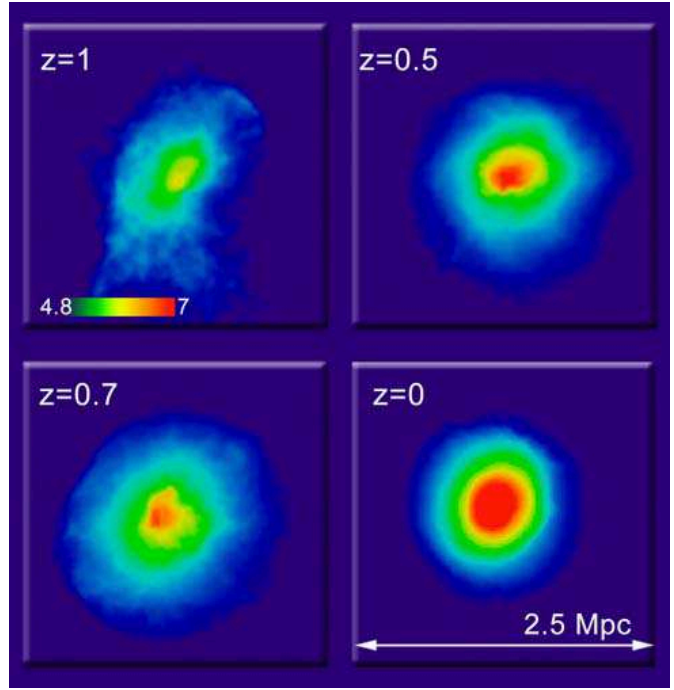}} \caption{X-ray
surface brightness maps for different redshifts (z=1, 0.7, 0.5 and
0) for  model cluster B. The maps have 2.5 Mpc comoving on a side
and give the x-ray emission in a logarithmic scale.}
\label{xsb_maps604}
\end{center}
\end{figure}

In Fig. \ref{xsb_maps604} the X-ray surface brightness maps for
model cluster B at different redshifts (z=1, 0.7, 0.5 and 0) are
presented. In contrast to model cluster A, this cluster shows a
rather relaxed shape already at z=0.7.

\begin{figure}
\begin{center}
{\includegraphics[width=\columnwidth]{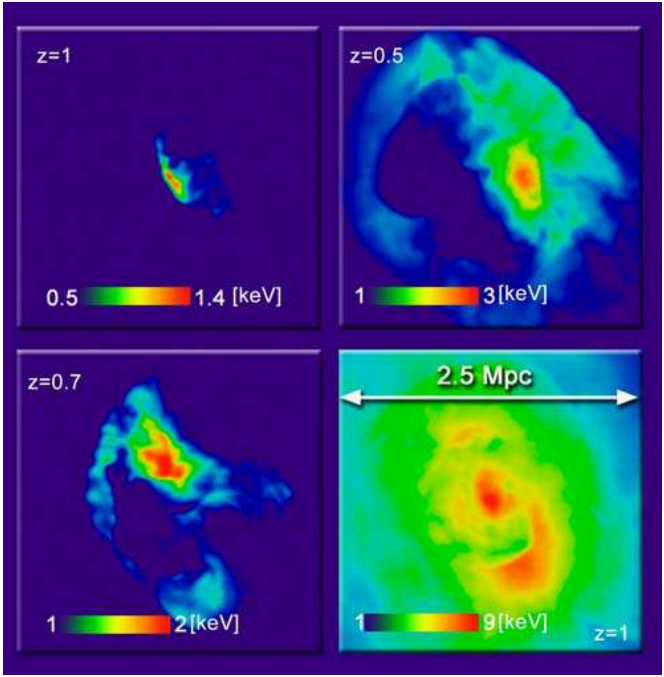}} \caption{X-ray
weighted temperature maps for different time steps (z=1, 0.7, 0.5
and 0) for model cluster A. The maps have a size of 2.5 Mpc comoving
and give the temperature in keV linear scale.} \label{xt_maps}
\end{center}
\end{figure}

\begin{figure}
\begin{center}
{\includegraphics[width=\columnwidth]{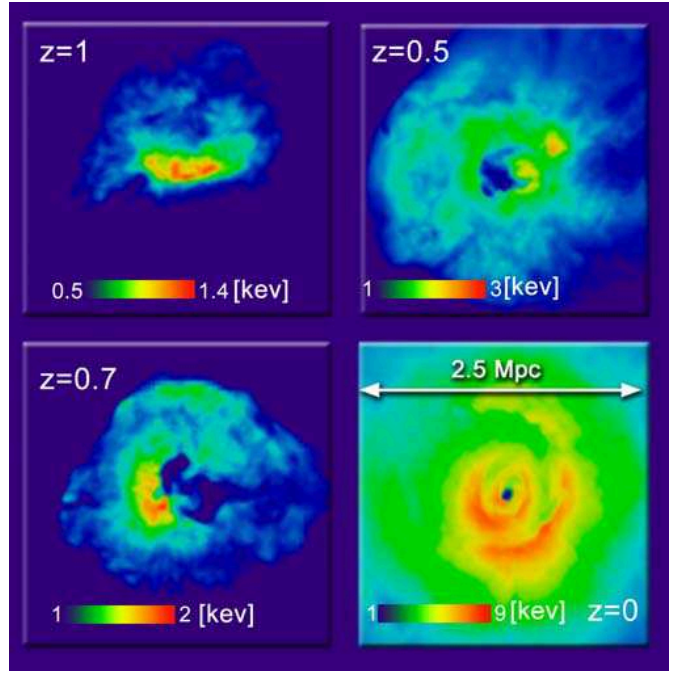}} \caption{X-ray
weighted temperature maps for different time steps (z=1, 0.7, 0.5
and 0) for model cluster B. The maps have a size of 2.5 Mpc comoving
and give the temperature in keV linear scale.} \label{xt_maps_604}
\end{center}
\end{figure}

In Figs. \ref{xt_maps} and \ref{xt_maps_604} we show X-ray weighted
temperature maps for the model clusters at the same redshifts as in
Fig. \ref{xsb_maps602} and Fig. \ref{xsb_maps604}. The hierarchical
growth of the model cluster A, which is a direct consequence of the
initial power spectrum, is visible. After the formation along a
filamentary structure, the cluster builds up via several mergers,
which leads to the complex dynamical state of the gas. At the
present epoch the cluster has a distinct temperature distribution,
see the 2.5 Mpc comoving maps, with a drop towards the cluster
outskirts in the order of 50\%. Model cluster B shows a common
feature of relaxed clusters, namely a drop in  temperature towards
the innermost region, i.e. a cool core. As we are not interested in
the gas dynamics of the cooling flow region, but rather in the
large-scale distribution of the metals in the ICM, we do not
introduce a heating source to counterbalance the overcooling.
However, due to our limited spatial resolution we underestimate the
cooling in this region, avoiding a cooling catastrophe. Hence,
except of the innermost computational cells, the radial temperature
profile resembles that of observed relaxed galaxy clusters.

\begin{figure}
\begin{center}
{\includegraphics[width=\columnwidth]{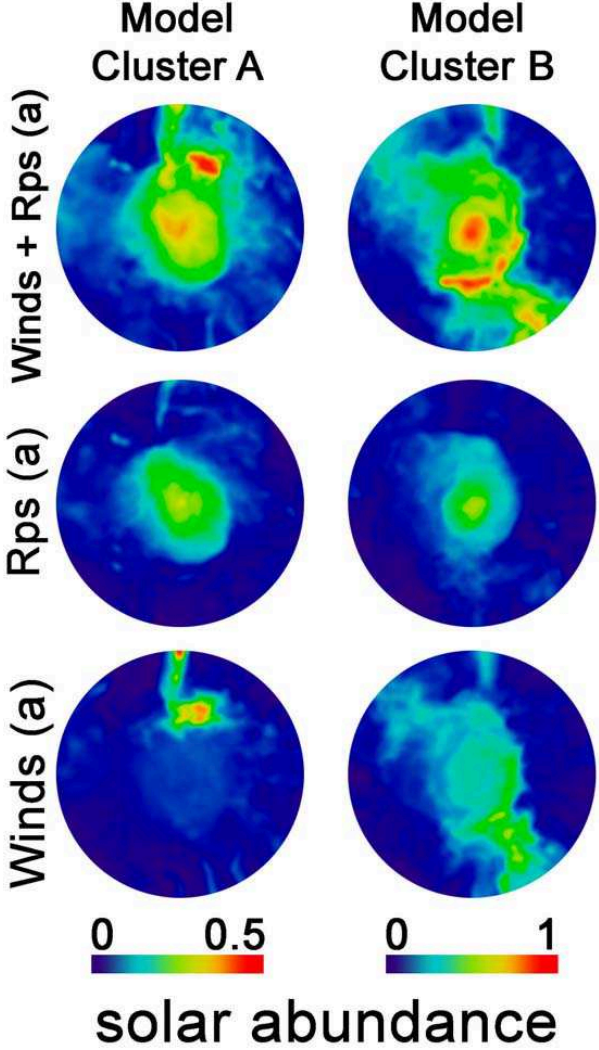}} \caption{X-ray
weighted metal maps for the two model clusters in an area with 2000
kpc diameter for different enrichment processes at redshift 0,
enrichment process combination $a$.} \label{metal_maps_old}
\end{center}
\end{figure}

\begin{figure}
\begin{center}
{\includegraphics[width=\columnwidth]{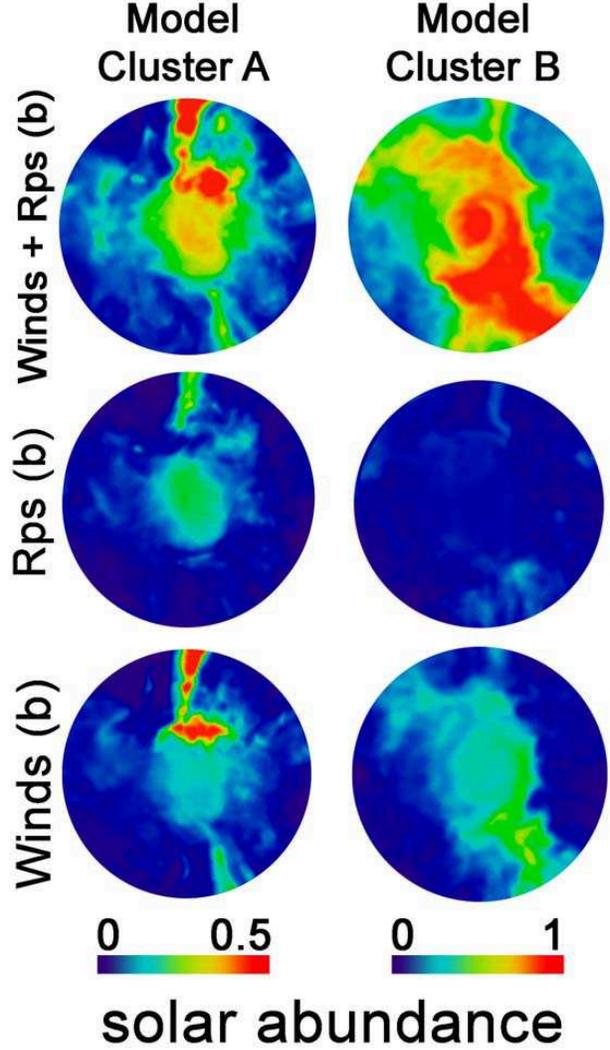}} \caption{X-ray
weighted metal maps for the two model clusters in an area with 2000
kpc diameter for different enrichment processes at redshift 0,
enrichment process combination $b$.} \label{metal_maps_new}
\end{center}
\end{figure}

In order to investigate the strength of the different enrichment
processes and the spatial distribution of the metals, we construct
X-ray weighted metal maps from our 3D data. These maps can be
compared to observations. In Fig. \ref{metal_maps_old} and Fig.
\ref{metal_maps_new} metallicity maps of the two model clusters for
the different enrichment process combinations descriptions $a$ and
$b$ are presented. We construct them for redshift z=0 within an area
of 2 Mpc in diameter. The left column corresponds to the merging
model cluster A, whereas the right column shows the relaxed system
of model cluster B. In order to highlight the different spatial
distribution of the metals emerging from the two different
enrichment processes, i.e. ram-pressure stripping and galactic
winds, we show separate metal maps for each enrichment process along
with a map that gives the overall metal distribution.

\subsection{Mass loss by ram-pressure stripping and galactic winds}

To address the question on the different efficiencies of the two
enrichment processes, we give in Table \ref{Mass-losses-winds} and
Table \ref{Mass-losses-rps} the integrated mass loss for the cluster
galaxies in two different redshift regimes for ram-pressure
stripping and galactic winds for enrichment process combination $a$
and $b$. As the clusters form at redshifts z$<$2 and because
ram-pressure stripping depends highly on the density of the
environment, it is more important in this redshift regime. In the
case of enrichment combination $b$ and model cluster B ram-pressure
stripping is less important for enriching the ICM. As galactic winds
are not suppressed by ram-pressure stripping or the pressure of the
ICM they nearly exhaust the gas reservoir of the galaxies. Galactic
winds, on the contrary, induce higher mass losses at higher
redshifts. The mass loss by this process is mainly driven by the
supernova rate of a system, which is tightly linked to the star
formation rate. As the star formation rate peaks at redshifts larger
than 2 (e.g. Hopkins 2004), the mass loss of galactic winds is the
more dominant process for the removal of gas from galaxies in this
redshift regime. Ram-pressure stripping is in terms of total mass
loss a factor of three more efficient than winds in model cluster A
and 50\% in model cluster B for the enrichment process combination
$a$. This can be explained by the merger events in model cluster A,
which introduce disturbances in the ICM. These increase the ram
pressure onto the cluster galaxies. In model cluster B the
difference of mass loss between ram-pressure stripping and galactic
winds is lower compared to model cluster A. Compared to the merging
system, where also in the outer regions of the cluster galaxies are
affected by ram-pressure, in relaxed systems ram-pressure is
predominately present in the central cluster regions and therefore
not so efficient. In the case of enrichment process combination $b$
the ratio of the mass loss between ram-pressure stripping and
galactic winds is comparable to enrichment process combination $a$
for model cluster A. Model cluster B shows a different picture,
mainly because enrichment process combination $b$ does deplete
remove too much gas from the galaxies by galactic winds, leading to
a small amount gas left in the galaxies at the epochs when
ram-pressure stripping becomes more important. It seems that in this
case we underestimate the interaction between ram-pressure stripping
and galactic winds.

\begin{table}
\caption[]{Fractions of the integrated mass loss due to galactic
winds for model cluster A and B.}
\begin{tabular}{c | c c c c c }
\hline \hline Model & redshift  & Winds $a$ & Winds $b$\cr \hline  A
& z$>$2 & 20.5\% & 13.5\%\cr A & z$<$2 & 79.5\% & 87.5\%\cr   A &
wind mass loss & 1.2$\times 10^{11}$ M$_{\odot}$ & 2.5$\times
10^{11}$ M$_{\odot}$\cr  \hline B & z$>$2 & 21.4\% & 12.3\%\cr B &
z$<$2 & 78.6\% & 87.7\%\cr B & wind mass loss & 4.4$\times 10^{11}$
M$_{\odot}$ & 1.7$\times 10^{12}$ M$_{\odot}$\cr \hline
\end{tabular}
\label{Mass-losses-winds}
\newline
Total mass loss (galactic winds and ram-pressure stripping) model
cluster A enrichment combination $a$: 4.8$\times 10^{11}$
M$_{\odot}$, model cluster B: 1.12$\times 10^{12}$ M$_{\odot}$ -
enrichment combination $b$: 5.6$\times 10^{11}$ M$_{\odot}$, model
cluster B: 1.9$\times10^{12}$ M$_{\odot}$ in a (5 Mpc)$^3$ comoving volume.\\
\end{table}

\begin{table}
\caption[]{Fractions of the integrated mass loss due to ram-pressure
stripping for model cluster A and B.}
\begin{tabular}{c | c c c c c}
\hline \hline Model & redshift  & RPS $a$ & RPS $b$\cr \hline  A &
z$>$2 & 2.7\% & 4.7\%\cr A & z$<$2 & 97.3\% & 95.3\%\cr   A & rps
mass loss & 3.6$\times 10^{11}$ M$_{\odot}$ & 3.1$\times 10^{11}$
M$_{\odot}$ \cr  \hline B & z$>$2 & 4.8\% & 42.4\%\cr B & z$<$2 &
95.2\% & 57.6\%\cr B & rps mass loss & 6.8$\times 10^{11}$
M$_{\odot}$ & 1.7$\times 10^{11}$ M$_{\odot}$\cr \hline
\end{tabular}
\label{Mass-losses-rps}
\newline
Total mass loss (galactic winds and ram-pressure stripping) model
cluster A enrichment combination $a$: 4.8$\times 10^{11}$
M$_{\odot}$, model cluster B: 1.12$\times 10^{12}$ M$_{\odot}$ -
enrichment combination $b$: 5.6$\times 10^{11}$ M$_{\odot}$, model
cluster B: 1.9$\times10^{12}$ M$_{\odot}$ in a (5 Mpc)$^3$ comoving volume.\\
RPS....Ram-Pressure Stripping
\end{table}

\begin{figure}
\begin{center}
{\includegraphics[width=\columnwidth]{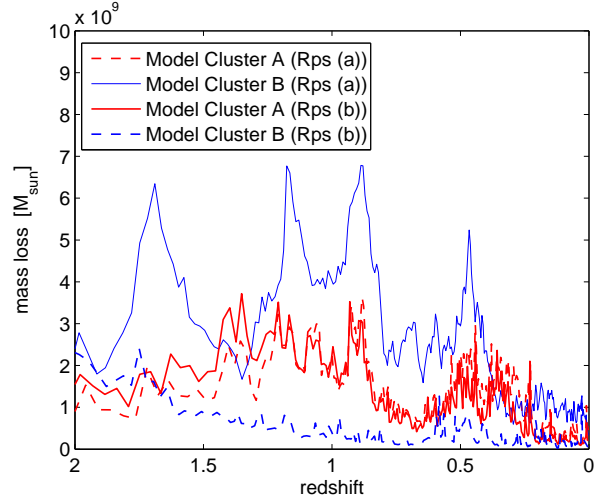}} \caption{Mass
loss due to ram-pressure stripping as a function of redshift for the
two model clusters, in a 5 Mpc on a side cube around the cluster
centre, for two different enrichment process descriptions.}
\label{outflow_rps}
\end{center}
\end{figure}

In Fig. \ref{outflow_rps} the mass loss of all galaxies due to
ram-pressure stripping in a (5 Mpc)$^3$ cube for both enrichment
process description is presented. As ram-pressure stripping is
strongly dependent on the environment the mass loss increases as the
cluster forms, i.e. in the redshift regime z$<$2. In addition
merging events increase the density in the ICM temporarily, leading
to strong mass losses of galaxies passing this regions.
Additionally, groups of galaxies which belong to sub-clusters and
enter the main cluster, fall into the cluster with high velocities
and therefore increase the ram pressure. This again leads to
temporarily high mass losses. Although the massive model cluster B
show no major merger events, ram-pressure stripping is even more
efficient than in the merging model cluster A for the enrichment
process combination $a$. Due to the higher mass of model cluster B
(more than twice as massive than cluster A), more massive galaxies
which harbour more gas mass in the discs reside in it. As in
addition the density in the cluster's central regions is higher the
mass loss is larger in model cluster B.

\begin{figure}
\begin{center}
{\includegraphics[width=\columnwidth]{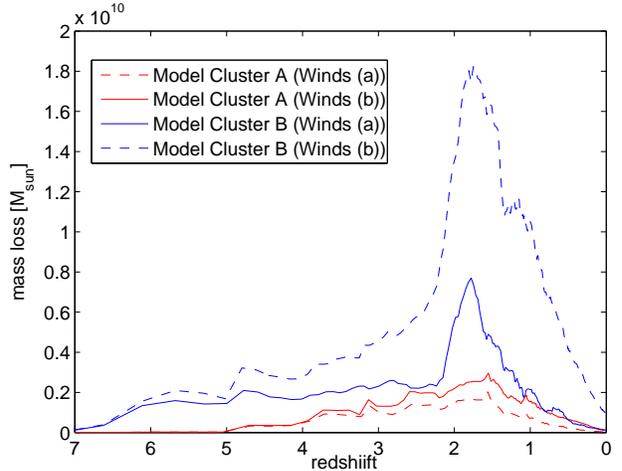}} \caption{Mass
loss due to galactic winds as a function of redshift for the two
model clusters, in a cube around the cluster centre with 5 Mpc on a
side, for two different enrichment process descriptions. Note that
enrichment process description $b$ has higher mass losses, as there
is no feedback of ram-pressure stripping on the galactic winds.}
 \label{outflow_wind}
\end{center}
\end{figure}

Fig. \ref{outflow_wind} gives the mass loss of all galaxies in a (5
Mpc)$^3$ cube for the galaxies for both enrichment process
combinations due to galactic wind. The increase of mass loss by
galactic winds at redshift z$\sim$2 in model cluster B has two
reasons. There is an obvious connection between the mass loss
efficiency of galactic winds and ram-pressure stripping and the
formation history of a galaxy cluster. If the system is a merging
system (several major events at all epochs) then the differences in
mass loss between galactic winds and ram-pressure are in the order
of several 100\%, resulting in ram-pressure stripping as the
dominant enrichment process for both enrichment process
combinations. In relaxed, massive systems the ratio drops
significantly, for model cluster B ram-pressure stripping is only
50\% more efficient for enrichment process combination $a$. The
enrichment combination $b$ leads to a contrary picture. In this case
the mass loss by galactic winds is an order of magnitude higher than
due to ram-pressure stripping. The amount of matter lost by
cluster-galaxies is a strong function of the number of cluster
galaxies. As a more massive system has more galaxies the absolute
value of mass loss is higher than in the merging system. To what
extent these mass losses enrich the ICM will be discussed in the
next section.

\subsection{Metallicity profiles}

\begin{figure}
\begin{center}
{\includegraphics[width=\columnwidth]{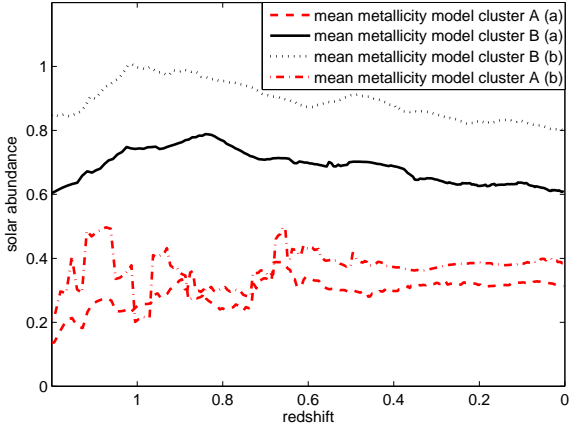}}
\caption{Evolution of the mean metallicity for both model clusters
taken from the X-ray emission within a circle of 500 kpc radius for
both enrichment process combinations $a$ and $b$.}
\label{metal_evo_winds_rps}
\end{center}
\end{figure}

Observers often give metallicity profiles of galaxy clusters. By
integrating X-ray events in radial bins it is possible to construct
X-ray spectra to investigate metal abundances in a given annulus.
With a longer exposure time or for a higher luminosity of the galaxy
cluster, more events by X-ray photons arrive at the detector. This
makes it possible to increase the resolution of the metallicity
profiles or to construct 2D maps of the metallicity of the observed
galaxy cluster. To compare our simulated ICM properties with
observations we provide metal profiles and X-ray weighted
metallicity maps.\\
The evolution of the mean metallicity in an area with 500 kpc
physical radius around the cluster centre is shown for both model
clusters and enrichment process combinations in Fig.
\ref{metal_evo_winds_rps}. At z$<$1.2 there is a clear difference in
metallicity present, which belongs to the different formation epoch
for the both clusters. The relaxed system has already assembled,
leading to higher metallicities at z=1.2. As metallicity is a
measure of heavy elements to primordial the merging system, building
up by mergers introduces enriched ICM to the cluster region at all
epochs, leading to a slightly increase in the mean metallicity.
Model cluster B is has nearly constant mean metallicities in the
redshift regime z$<$1.2. The difference is given by the different
amount of primordial ICM mass accretion in the same interval.
Whereas the enrichment combinations $a$ and $b$ deliver nearly the
same results for model cluster A, there is a difference for model
cluster B. In this case nearly all metals are lost by galactic
winds, as they are not suppressed by any physical process like
pressure of the ICM onto the galaxies (Schindler et al. 2005).

\begin{figure}
\begin{center}
{\includegraphics[width=\columnwidth]{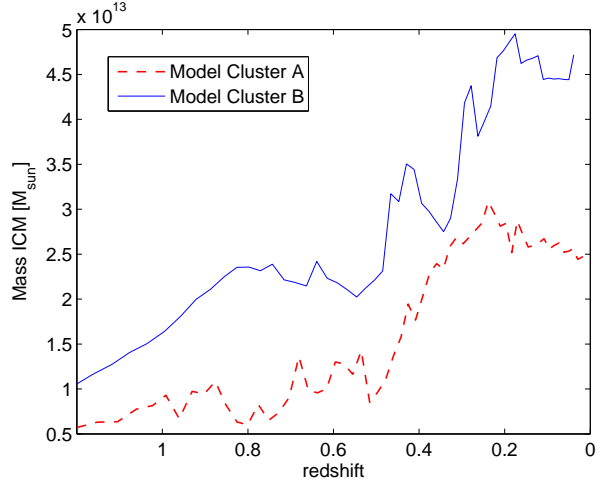}}
\caption{Evolution of the ICM mass in a sphere with 2 Mpc radius
around the cluster centre.} \label{mass_ICM_evo}
\end{center}
\end{figure}

Fig. \ref{mass_ICM_evo} shows the evolution of the ICM mass within a
sphere with a radius of 2 Mpc comoving around the cluster centre.
Besides the strong signs of sub-clusters and groups falling into the
sphere, the main difference between model cluster A and B is the
growth of mass between z=1.2 and z=0. Metals ejected at higher
redshift and falling towards the cluster centre have to mix with yet
unpolluted gas, resulting in a decrease of metallicity in model
cluster B. In model cluster A the amount of ICM in the same redshift
interval does not increase so much, despite during merger events.
Therefore the metals ejected by galactic winds and removed by
ram-pressure stripping do increase the overall metallicity in the
500 kpc radius area around the cluster centre. This slight increase
is consistent with observations, e.g. Balestra et al. (2006). In the
case of the massive model cluster B this seems not to be the case.
Combining different clusters with different luminosities might
explain this discrepancy.

\begin{figure}
\begin{center}
{\includegraphics[width=\columnwidth]{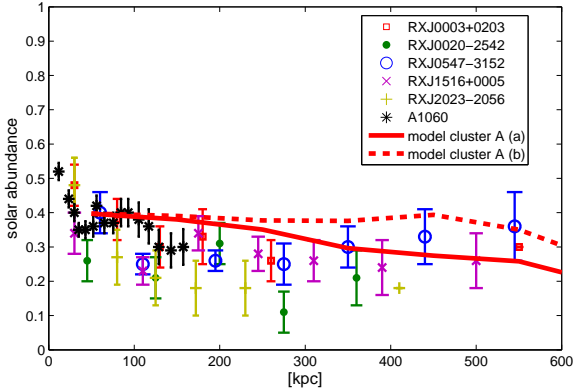}}
\caption{Metallicity profiles for six non-cooling flow clusters and
model cluster A for both enrichment process combinations $a$ and
$b$.}
 \label{cluster_profile}
\end{center}
\end{figure}

In Fig. \ref{cluster_profile} observed metallicity profiles for six
different non-cooling flow clusters are shown. The data points are
taken from Pratt et al. (2006) (RXJ0003+0203, RXJ0020-2542,
RXJ0547-3152, RXJ1516+0005 and RXJ2023-2056) and Hayakawa et al.
(2006) (A1060), as they present non-cooling flow clusters. A common
striking feature for all measured radial metallicity profiles is a
nearly flat distribution within a radius of 600 kpc around the
cluster centre. In contrast to cooling flow clusters, which show an
enhancement of metallicities towards the cluster centre, e.g. De
Grandi et al. (2004), non-cooling flow clusters have typical mean
metallicities in the order of 0.3 - 0.5 in solar abundance in the
central region. We include in Fig. \ref{cluster_profile} the
metallicity profiles for model cluster A with both enrichment
process combinations $a$ and $b$, see e.g. Fig. \ref{metal_maps}. We
are able to reproduce both the observed spatial distribution and the
observed amount of metals when taking into account both enrichment
processes. Due to the limited numerical resolution, we are not able
to follow the increase in the observed profiles in the very central
($\sim$ 50 kpc) region of the clusters. By taking into account two
enrichment processes, namely galactic winds and ram-pressure
stripping, the profiles show the relatively flat behaviour over
large radii.

\begin{figure}
\begin{center}
{\includegraphics[width=\columnwidth]{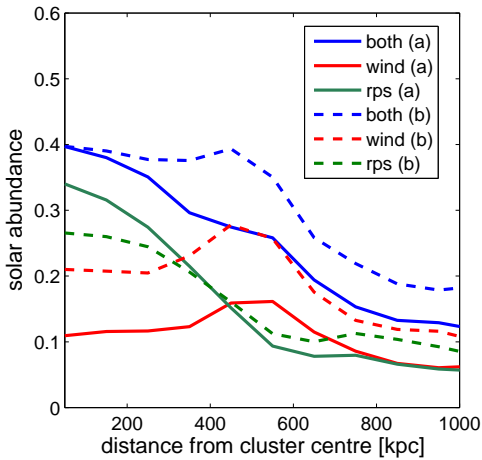}}
\caption{Metallicity profiles of the model clusters, presented in
Fig. \ref{metal_maps_old} and \ref{metal_maps_new}. The profiles for
the metals ejected by galactic winds, ram-pressure stripping and
both enrichment precesses together for the two enrichment process
combinations $a$ and $b$ are shown.} \label{profiles_sim_602}
\end{center}
\end{figure}

In Fig. \ref{profiles_sim_602} the metallicity profiles for the
metals, lost by galactic winds and by ram-pressure stripping in
model cluster A for both enrichment processes combinations $a$ and
$b$ are presented. Galactic winds are present at all radii,
ram-pressure stripping is more concentrated towards the centre. This
would in principle lead to different element ratios at radii below
and above $\sim$500 kpc, which should be present in very long
exposure X-ray observations.

\begin{figure}
\begin{center}
{\includegraphics[width=\columnwidth]{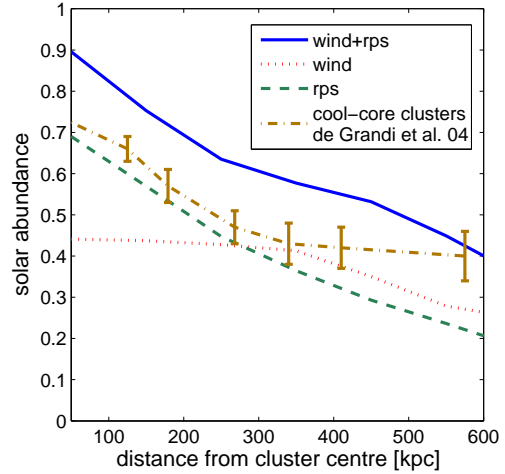}}
\caption{Metallicity profiles of the model clusters, presented in
Fig. \ref{metal_maps_old}. The profiles for the metals ejected by
galactic winds, ram-pressure stripping and both enrichment processes
for enrichment process combinations $a$ are shown. The dashed-dotted
line corresponds to the mean of cool-core clusters presented in De
Grandi et al. (2004).} \label{profiles_sim_604}
\end{center}
\end{figure}

Fig. \ref{profiles_sim_604} shows the different profiles
(ram-pressure stripping, galactic winds and both) of model cluster B
for enrichment process combination $a$ as well as a mean profile for
cool-core clusters from De Grandi et al. (2004). In the case of
massive clusters, galactic winds are more efficient at radii grater
than 300 kpc, whereas ram-pressure stripping enriches the cluster
centre more efficiently. The flat distribution present in the
non-cooling flow systems alters to a distribution clearly increasing
towards the centre. The simulation overestimates the observed
metallicity. The slope of the observed iron profile is reproduced by
the simulation out to 400 kpc. Above 400 kpc the slope is too steep,
compared to observations. The enrichment process combination $b$
gives a different result, leading to very small mass loss by
ram-pressure stripping.

\section{Discussion and conclusions}

We investigate the efficiency of galactic winds and ram-pressure
stripping in combined N-body/hydrodynamic simulations of the
intra-cluster medium with semi-numerical galaxy modelling. The
terminus mass loss in our simulations refers to galactic material,
which is not gravitationally bound to a galaxy anymore. From the
simulations we calculate X-ray weighted metallicity maps and
profiles as well as mass loss histories for the model cluster
galaxies. The profiles are compared with observations. A major
uncertainty are the metallicities of the ejecta and stripped gas of
the galaxies. This leads in our setup to higher or lower overall
metallicities in the ICM, but would not affect the spatial
distribution of metals in the ICM. Given the model assumptions, we
find the following

\begin{enumerate}

\item We confirm the inhomogeneous distribution of heavy elements found in the
intra-cluster medium.\\

\item The mass loss due to galactic winds is half
the mass loss caused by ram-pressure stripping in the case of a
merging galaxy cluster if ram-pressure stripping is allowed to
suppress the outflow by galactic winds. In the case of no influence
of ram-pressure stripping onto galactic winds the mass loss by
galactic winds is 80\% of the mass loss by ram-pressure stripping.
In the case of a relaxed, massive system the mass loss by
ram-pressure stripping is three times higher than the mass loss by
galactic winds if ram-pressure stripping suppresses the outflow of
galactic winds. In this case ram-pressure stripping is the
dominating enrichment process in cool-core galaxy clusters. If
galactic winds are not affected by ram-pressure stripping the
efficiency of both processes permute, leading to nearly no mass loss
by ram-pressure stripping.\\
For merging galaxy clusters the number of galaxies losing mass by
ram-pressure stripping is nearly as high as in a relaxed system.
This difference can be explained by sub-cluster merger events,
increasing the density of the ICM and the
relative velocities of the galaxies with respect to the ICM.\\

\item Whereas galactic winds trigger most of the mass loss in the
redshift regime z$>$2, ram-pressure stripping increases the mass
loss in the redshift interval z$<$2, in which the clusters
start to form. \\

\item We traced the evolution of the metallicity in the ICM and found
an increase of the metallicity in a 1 Mpc area around the cluster
centre in the redshift regime z$<$1.2 for the merging system. The
relaxed, massive galaxy cluster shows nearly a constant metallicity
in the same redshift regime. Unpolluted ICM entering the cluster
region and the different efficiency of the enrichment processes as a
function of redshift explains this behaviour. In the case of the
merging system enrichment processes are relatively more efficient
over all redshifts, leading to a slight increase for the mean
metallicity. In the relaxed system ram-pressure stripping and
galactic winds do not enrich unpolluted ICM to such an amount, that
the metallicity of the
whole cluster increases.\\

\item Comparing the metallicity profiles of observed merging
clusters with model clusters, we find good agreement in the spatial
distribution of the metallicity for merging galaxy clusters. In the
case of relaxed, massive systems the slope of the observed
metallicity profile is well reproduced in the central $\sim$300 kpc,
whereas beyond $\sim$300 kpc the observed profiles are shallower.\\

\item A comparison of the metals lost by galactic winds and
ram-pressure stripping leads to the conclusion that at radii less
than 500 kpc matter lost by ram-pressure stripping dominates the
profiles, whereas in the outer regions galactic winds contribute to
the same amount. If galactic winds are not affected by ram-pressure
stripping, galactic winds are more efficient over the whole galaxy
cluster in the case of a massive system. If distinct metal blobs
caused by galaxies with strong galactic winds fall towards the
cluster centre, humps in the metallicity profiles are visible (see
Figs. \ref{metal_maps_old}, \ref{metal_maps_new} and
\ref{profiles_sim_602}).\\

\end{enumerate}

\section*{Acknowledgements}

The authors are grateful to the anonymous referee for his/her
fruitful comments that helped to improve the paper. The authors
would like to thank Volker Springel for providing GADGET2 and Paul
Ricker for clarifying explanations. Edmund Bertschinger and Rien van
de Weygaert are acknowledged for providing their constrained random
field code.  The authors acknowledge the Austrian Science Foundation
(FWF) through grant P18523-N16 and grant P18416-N16, the German
Science Foundation (DFG) through grant Zi 663/6-1. Thomas Kronberger
is a recipient of a DOC-fellowship of the Austrian Academy of
Science. Magdalena Mair is a recipient of a Doktoratsstipendium der
Universit\"at Innsbruck and a fellowship of Mils Electronic. In
addition, the authors acknowledge the ESO Mobilit\"atsstipendien des
bm:bwk (Austria), the Tiroler Wissenschaftsfonds (Gef\"ordert aus
Mitteln des vom Land Tirol eingerichteten Wissenschaftsfonds), the
UniInfrastrukturprogramm 2005/06.

\end{document}